\documentclass{article}

\usepackage{PRIMEarxiv}

\usepackage[utf8]{inputenc} 
\usepackage[T1]{fontenc}    
\usepackage[hidelinks]{hyperref}  
\usepackage{hyperref}       
\usepackage{url}            
\usepackage{booktabs}       
\usepackage{amsfonts}       
\usepackage{amsmath}
\usepackage{latexsym}
\usepackage{gensymb}
\usepackage{arydshln}
\usepackage{nicefrac}       
\usepackage{microtype}      
\usepackage{lipsum}
\usepackage{fancyhdr}       
\usepackage{graphicx}       
\graphicspath{{media/}}     

\usepackage{pdflscape}
\usepackage{afterpage}
\usepackage{colortbl}
\usepackage{makecell}
\usepackage{multirow}
\usepackage{subcaption}

\newcommand\blfootnote[1]{%
  \begingroup
  \renewcommand\thefootnote{}\footnote{#1}%
  \addtocounter{footnote}{-1}%
  \endgroup
}
  
\title{Uncertainty-Aware Multiple-Instance Learning for Reliable Classification: Application to
Optical Coherence Tomography
}

\author{
  Coen de Vente\normalfont{\textsuperscript{a,b,c,*}} \\
   \And
  Bram van Ginneken\normalfont{\textsuperscript{c}} \\
  \AND
  Carel B. Hoyng\normalfont{\textsuperscript{d}} \\
  \And
  Caroline C. W. Klaver\normalfont{\textsuperscript{d,e}} \\
  \\ \\
  \And
  Clara I. S\'{a}nchez\normalfont{\textsuperscript{a,b}} \\
}

\begin{document}
\maketitle

\begin{abstract}
Deep learning classification models for medical image analysis often perform well on data from scanners that were used during training. However, when these models are applied to data from different vendors, their performance tends to drop substantially. Artifacts that only occur within scans from specific scanners are major causes of this poor generalizability.
We aimed to improve the reliability of deep learning classification models by proposing Uncertainty-Based Instance eXclusion (UBIX). This technique, based on multiple-instance learning, reduces the effect of corrupted instances on the bag-classification by seamlessly integrating out-of-distribution (OOD) instance detection during inference.
Although UBIX is generally applicable to different medical images and diverse classification tasks, we focused on staging of age-related macular degeneration in optical coherence tomography.
After being trained using images from one vendor, UBIX showed a reliable behavior, with a slight decrease in performance (a decrease of the quadratic weighted kappa ($\kappa_w$) from 0.861 to 0.708), when applied to images from different vendors containing artifacts; while a state-of-the-art 3D neural network suffered from a significant detriment of performance ($\kappa_w$ from 0.852 to 0.084) on the same test set. 
We showed that instances with unseen artifacts can be identified with OOD detection and their contribution to the bag-level predictions can be reduced, improving reliability without the need for retraining on new data. This potentially increases the applicability of artificial intelligence models to data from other scanners than the ones for which they were developed.

\end{abstract}


\blfootnote{\textsuperscript{a}Quantitative Healthcare Analysis (QurAI) Group, Informatics Institute, University of Amsterdam, Amsterdam, Noord-Holland, Netherlands}
\blfootnote{\textsuperscript{b}Department of Biomedical Engineering and Physics, Amsterdam University Medical Center, Amsterdam, Noord-Holland, Netherlands}
\blfootnote{\textsuperscript{c}Diagnostic Image Analysis Group (DIAG), Department of Radiology and Nuclear Medicine, Radboudumc, Nijmegen, Gelderland, Netherlands}
\blfootnote{\textsuperscript{d}Department of Ophthalmology, Radboudumc, Nijmegen, Gelderland, Netherlands}
\blfootnote{\textsuperscript{e}Ophthalmology \& Epidemiology, Erasmus MC, Rotterdam, Zuid-Holland, Netherlands}
\blfootnote{*Corresponding author: \texttt{research@coendevente.com} (Coen de Vente)}

\section{Introduction}
\interfootnotelinepenalty=10000

\begin{figure*}[ht]
    \centering
    \includegraphics[width=.9\textwidth]{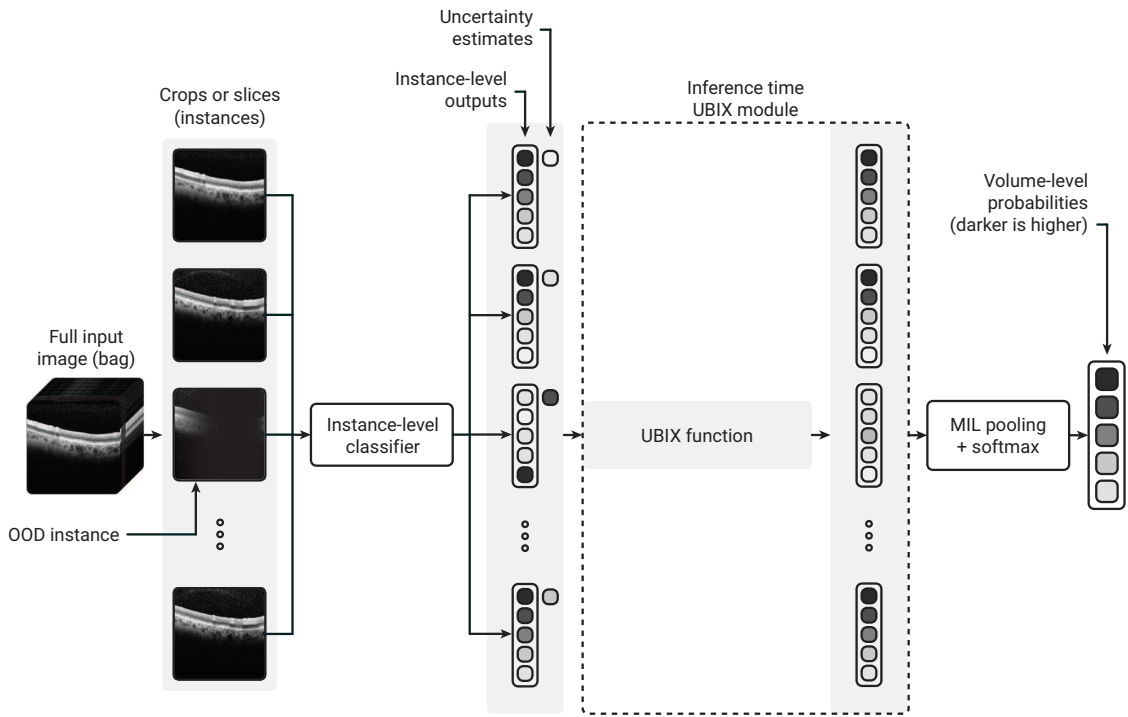}
    \caption{Overview of the method. Each MIL instance is fed into the same classifier. During inference time, a UBIX function converts pre-UBIX instance-level logits based on their respective uncertainties to post-UBIX logits, which are reduced for OOD instances. The instance-level post-UBIX logits are then converted to bag-level outputs using MIL pooling. During training, the pre-UBIX logits are fed directly into the MIL pooling function.}
    \label{fig:method_summary}
\end{figure*}

Deep learning models for medical image analysis applications are often trained on data that is acquired with one or a selected number of scanner types and/or acquisition protocols. When applying these trained models on data from different scanners or protocols, the performance tends to plummet (\cite{Yana20,Fauw18}). This negatively affects the reliability of these systems, which is a main aspect of trustworthy AI (\cite{Gonz21b,Euro19}), and its wide integration and adoption in clinical practice.
In general, convolutional neural networks (CNNs) are known to fail when they are applied under dataset shift or to out-of-distribution (OOD) datasets; and approaches to address this effect are being investigated (\mbox{\cite{Ovad19}}). This OOD nature of the data occasionally only stems from local areas in images, such as local artifacts. These local artifacts occur frequently in data from specific vendors or particular scanning protocols, and can be found in multiple medical imaging fields, such as contrast-enhanced mammography data (\cite{Nepp21}) and optical coherence tomography (\cite{Bazv20}). In images with these types of artifacts, there generally are sufficient parts in a sample which are in-distribution (ID) to form a correct prediction if the model would in some way only focus on those parts of the data and neglect the OOD areas.

To achieve this increased robustness to local OOD areas in images, we propose Uncertainty-based Instance eXclusion (UBIX). This approach builds upon multiple-instance learning (MIL), a form of weakly supervised learning popular in medical image analysis (\cite{Chep19}, \cite{Ilse18}). In MIL, a labeled bag (usually the whole input image) consists of multiple unlabeled instances (usually image patches, regions or slices). During deep MIL, instances are considered individually by a neural network and the instance-level outputs are subsequently combined, equally contributing, to obtain a bag-level prediction using a MIL pooling function (\cite{Wang18}). Instead of assuming equal contribution, the UBIX approach assumes that some of the instances might be corrupted due to local artifacts, identifies these instances on-the-fly using uncertainty estimation, and reduces or ignores its contribution to the bag-level prediction using the so-called UBIX function before the MIL pooling function (see Fig. \ref{fig:method_summary}). To the best of our knowledge, this is the first method that uses OOD detection in such a manner to increase reliability.

Although our method is applicable to any instance definition (such as 2D or 3D patches) or bag definition, we focus on MIL problems in which slices are instances and full 3D volumes are bags. Specifically, we focus on classifying age-related macular degeneration (AMD) in optical coherence tomography (OCT), as there is a plurality of manufacturers and scanner versions in the field of OCT. \cite{Swan17} list fourteen companies that produce OCT scanners for ophthalmic applications and the type of imaging artifacts that occur can differ substantially across scanners (\cite{Bazv20}). These artifacts include slices (B-scans) that are fully black due to blinking, vertically flipped B-scans, shadows and noise. For example, blinking artifacts are much less common in certain scanners, specifically ones with higher speed and eye tracking software (\cite{Bazv20}).

We evaluate the generalizability of our proposed models by training on data acquired with a scanner from one vendor, while evaluating with data from scanners of other vendors. We show that UBIX increases this generalizability using an ablation study. Moreover, we systematically analyze the ability of UBIX to detect OOD instances by gradually introducing artificial image artifacts that occur naturally as well. The trained algorithm is publicly available for inference on the online platform of Grand Challenge\footnote{\url{https://grand-challenge.org/algorithms/amd-classification-in-oct-with-ubix/}}.

\section{Related work}
\subsection{Multiple-instance learning in medical imaging}
One of the most common medical application in which MIL is applied is histopathology (\cite{Xu19a}, \cite{Pati19}, \cite{Chik20}, \cite{Tomc18}), mainly because it is very labor-intensive and time-consuming to manually annotate entire whole slide images on instance-level. \cite{Ilse18} used an attention-based MIL pooling layer and evaluated it on an MNIST-based dataset and histopathology datasets. Other medical modalities to which MIL with deep learning has been applied include ultrasound (\cite{Yin19,Shin18}), computed tomography (\cite{Han20a,Xu20b}) and magnetic resonance imaging (\cite{Zhu21,Qiu21}). Unlike our proposed method, these approaches suffer of low reliability when transferred to other distributions.

\subsection{Out-of-distribution detection}
OOD detection is the identification of samples that originate from a different distribution than the training distribution. Such samples generally have high model predictive uncertainties, given a good uncertainty estimation method. \cite{Hend16} proposed a simple baseline for OOD detection using the maximum class probability as confidence scores. Another early work was Monte Carlo dropout (MC-DO), in which they leveraged dropout to estimate uncertainty \cite{Gal16}. \cite{Ovad19} compared a number of methods for OOD detection and uncertainty estimation including MC-DO. They found that deep ensembling (\cite{Laks17}) was one of the top-performing methods for OOD detection. Since then, other popular methods for uncertainty estimation and OOD detection have been published (\cite{Hsu20,Liu20,Tack20}).

Uncertainty estimation has been investigated for medical image analysis as well, such as \cite{Mehr20}, who used deep ensembles to calibrate probabilities in segmentation maps. \cite{Call19} detected incorrect orientation or anatomy in X-rays using an OOD detection metric called FRODO, defined as the Mahalanobis distance of test samples to samples in the train set. Furthermore, \cite{Linm20,Linm23} used multi-head CNNs, an approach similar to deep ensembles, to detect images with lymphoma in histopathology as OOD samples.

In general, uncertainty estimation in medical images are used as an additional output to assess the behaviour of the developed models or to identify abnormalities as OOD samples. In contrast, our proposed approach takes into account OOD detection during inference to increase classification robustness against data shift.



\subsection{Robustness against data shift in OCT}
Related works have successfully applied machine learning methods for AMD classification from OCT, but did not specifically focus on robustness to OOD data (\cite{Apos17,Venh17a,Rast17,Kurm19,Lee17,Wang20c}). 




The following works studied robustness against data shift in OCT. 
\cite{Fauw18} used an OCT segmentation network of which the output was fed into a classification network, which in turn outputted a referral suggestion, diagnosis probabilities for multiple retinal disease features, such as choroidal neovascularization (CNV) and geographic atropy (GA), and volume estimations of drusen and epiretinal membranes. 
The error rate on their internal test set with OCTs from the same scanner, as the development set, i.e. Topcon, was 5.5\%, but the error rate increased to 46.6\% when transferred to an external set with OCTs from a different scanner, i.e. Heidelberg Spectralis. When retraining their segmentation network with data from this scanner, the error rate improved to 3.4\%. \cite{Seeb19} and \cite{Romo20} used a CycleGAN to transform OCT scans acquired on a device that was not used during training to have a similar appearance as the training data. For retinal fluid (\cite{Seeb19,Romo20}) and layer (\cite{Romo20}) segmentation, they observed a generalizability improvement when applying this domain adaptation technique, compared to traditional transformation strategies.


The main downside of these methods is their requirement for -- albeit annotated or not -- data from the new setting. Acquiring and annotating this new data, as well as any potential retraining, is a time-consuming and expensive process. Moreover, if these models are unknowingly applied in settings that are highly different from the development setting, models can fail silently, potentially causing misdiagnoses. We propose a method that reduces the performance drop when a model is transferred to a setting unlike its development setting, without the requirement for acquiring or labeling data originating from this new setting.



\section{Methods}
\begin{figure}[!t]
    \centering
    \includegraphics[width=.5\linewidth]{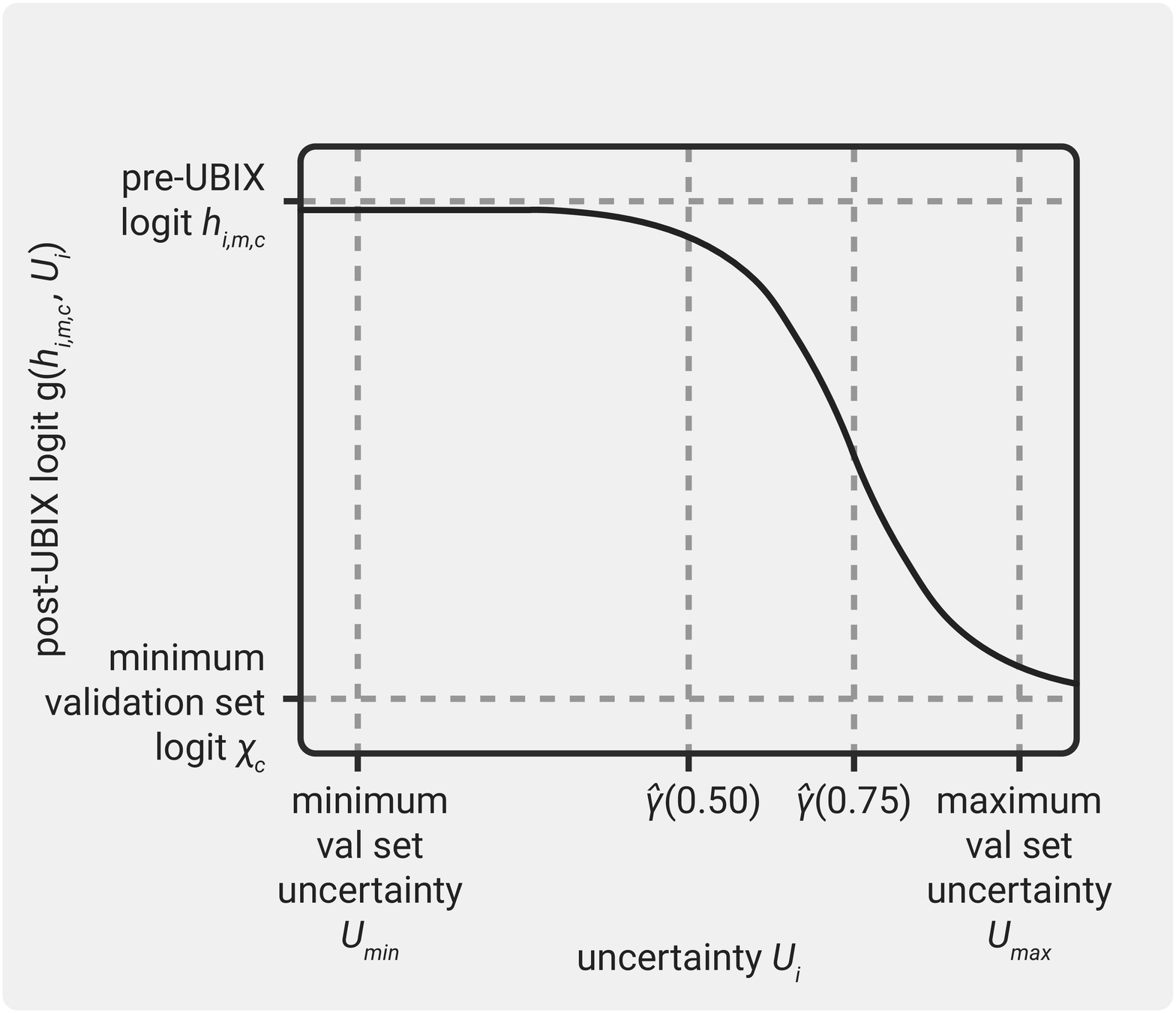}
    \caption{Plot of the UBIX function.}
    \label{fig:methods_ubix_function}
\end{figure}

\label{sec:methods}
In this section, we first introduce UBIX, a method that reduces the effect of corrupted instances on the bag classification by seamlessly integrating an OOD instance detection technique during inference (Section \ref{sec:methods_ubix}). The pipeline of the method is shown in Fig. \ref{fig:method_summary}. Subsequently, we introduce uncertainty estimation techniques used to identify OOD instances, including a novel ordinal uncertainty estimation technique, tailored to the classification of ordered classes (Section \ref{sec:uncertainty}).


\subsection{Uncertainty-based instance exclusion (UBIX)}
\label{sec:methods_ubix}
\subsubsection{UBIX during training time} 
Let $X=\{\mathbf{x}_1,\ldots,\mathbf{x}_I\}$ be a bag consisted of a set of instances $\mathbf{x}_i$, and $Y \in \{1,\ldots,C\}$ the bag label. We assume the number of instances in the bag, $I$, can vary between bags and each instance $\mathbf{x}_i$ in the bag also has an associated label $y_i \in \{1,\ldots,C\}$, which is not available. In this paper, we work with a staging problem, in which the classes are ordered, so we make the following assumption:


\begin{equation}
    Y = \max_i\{y_i\}.
    \label{eq:max_mil_assumption}
\end{equation}

We define an instance-level classifier which is an ensemble of neural networks, $f_{\theta_m}(\cdot)$ with parameters $\theta_m$ for each network in the ensemble $m \in \{1,\ldots,M\}$. Each network transforms instances $\mathbf{x}_i$ to logits $h_{i,m,c}$ for each class $1, \ldots, C$ such that $h_{i,m,c} \in \mathbb{R}$. Instance-level probabilities can be obtained with the softmax function:

\begin{equation}
p_{i,c}=\frac{1}{M}\sum_{m=1}^{M}p_{i,m,c}=\frac{1}{M}\sum_{m=1}^{M}\frac{e^{h_{i,m,c}}}{\sum_{d=1}^C{e^{h_{i,m,d}}}}.
\end{equation}

A MIL-pooling converts instance-level logits to bag-level logits. Following the assumption in Eq. \ref{eq:max_mil_assumption}, we define the MIL-pooling function as:

\begin{equation}
    z_{m,c} = \max_i\{h_{i,m,c}\},
\end{equation}

where $z_{m,c}$ is the bag-level logit for model $m$ in the ensemble and class $c$. The bag-level probability for class $c$ can be calculated using:

\begin{equation}
p_{c}=\frac{1}{M}\sum_{m=1}^{M}p_{m,c}=\frac{1}{M}\sum_{m=1}^{M}\frac{e^{z_{m,c}}}{\sum_{d=1}^C{e^{z_{m,d}}}}.
\end{equation}

\subsubsection{UBIX during inference time}
Each instance $x_i$ has an associated uncertainty estimate $U_i$. In Section \ref{sec:uncertainty}, we present various approaches of calculating this instance-level uncertainty. 

In UBIX, during inference, instance-level logits are modified using the UBIX function $g$ to consider the associated uncertainty:

\begin{equation}
g(h_{i,m,c},U_i) = \frac{h_{i,m,c} - \chi_c}{1 + e^{\delta(U_i - \hat\gamma)}} + \chi_c,
\end{equation}
where $\delta$ is a hyperparameter that determines the smoothness of $g$ and $\chi_c$ is:


\begin{equation}
\chi_c = \min_{m \in \{1,\ldots,M\}} \min_{x_i \in X_{val}} \{h_{i,m,c}\} ,
\end{equation}

where $X_{val}$ is the validation set.

Fig. \ref{fig:methods_ubix_function} shows a plot of this function. $\hat\gamma$ is the value for $h_{i,m,c}$ where the slope of $g$ is the steepest, defined as:

\begin{equation}
\hat{\gamma}(\gamma) = \gamma(U_{min} + U_{max}),
\end{equation}
where $\gamma \in [0,1]$ is a hyperparameter, $U_{min}$ is:

\begin{equation}
U_{min} = \min_{x_i \in X_{val}} \{U_i\},
\end{equation}
and:
\begin{equation}
U_{max} = \max_{x_i \in X_{val}} \{U_i\}.
\end{equation}


During inference, the MIL-pooling function takes as input $g(h_{i,m,c})$ instead of $h_{i,m,c}$:

\begin{equation}
    z_{m,c} = \max_i\{g(h_{i,m,c},U_i)\}.
\end{equation}

A special case of UBIX, where essentially the instances with an uncertainty exceeding a threshold $\tau$ are excluded from the bag, is the one for which $\delta = \infty$. The bag-level prediction is then calculated with this pruned bag. $\tau$ is a hyperparameter which can be optimized using the validation set. For this UBIX variant, the UBIX function is a step function.

\begin{table}[!t]
\centering
\caption{Equations for the different uncertainty measures used in this work. $U_i$ is the uncertainty for instance $x_i$, $C$ is the number of classes, $M$ is the number of models in the ensemble, and $p_{i,m,c}$ is the probability for instance $x_i$, in model $m$ of the ensemble, of class $c$.}
\begin{tabular}{@{}ll@{}}
\toprule
Uncertainty measure       & Equation \\ \midrule
\begin{tabular}[c]{@{}l@{}}Maximum class\\ probability\end{tabular} & $U_i = \max_{c=1}^C \frac{1}{M} \sum_{m=1}^M p_{i,m,c} $ \\ \midrule
\begin{tabular}[c]{@{}l@{}}Mean class\\ variance\end{tabular}       & \begin{tabular}[c]{@{}l@{}} $U_i = \frac{1}{C} \sum_{c=1}^C \frac{1}{M} \sum_{m=1}^M (p_{i,m,c} - \mu_{i,c})^2 $, \\ where $\mu_{i,c} = \frac{1}{M} \sum_{m=1}^M p_{i,m,c}$\end{tabular}  \\ \midrule
Ordinal variance          & \begin{tabular}[c]{@{}l@{}}$U_i = \frac{1}{M} \sum_{m=1}^M (q_{i,m} - \mu_{i})^2 $, where \\ $q_{i,m}=\sum_{c=1}^{C} (c-1) \cdot p_{i,m,c}$ and \\$\mu_{i} = \frac{1}{M} \sum_{m=1}^M q_{i,m}$ \end{tabular}        \\ \midrule
Entropy                   & \begin{tabular}[c]{@{}l@{}} $U_i = -\sum_{c=1}^C \mu_{i,x}\log{\mu_{i,c}} $, where \\ $\mu_{i,c} = \frac{1}{M} \sum_{m=1}^M p_{i,m,c}$\end{tabular}         \\ \midrule
Ordinal entropy           & \begin{tabular}[c]{@{}l@{}}
$U_i = -\sum_{c=1}^{C-1} (\sum_{d=1}^{c} \mu_{i,d} \log\sum_{d=1}^{c} \mu_{i,d}$ \\
$ + \sum_{d=c+1}^{C} \mu_{i,d} \log\sum_{d=c+1}^{C} \mu_{i,d})$, where \\ $\mu_{i,c} = \frac{1}{M} \sum_{m=1}^M p_{i,m,c}$
\end{tabular}         \\ \bottomrule
\end{tabular}
\label{table:uncertainty_measures}
\end{table}

\subsection{Instance-level uncertainty}
\label{sec:uncertainty}

The uncertainty $U_i$ of the instance $x_i$ in a bag $X$ is estimated using deep ensembling, as introduced by \cite{Laks17}.
We do not employ adversarial training, in contrast to what was done by \cite{Laks17}. 
Uncertainty estimates can be obtained from deep ensembles by combining the individual model outputs. This can be done in various ways, resulting in different uncertainty measures. Commonly used uncertainty measures are entropy (\cite{Ovad19,Nair20}), variance (\cite{Gal16,Nair20}) and maximum class probability (\cite{Hend16,Ovad19,Nair20}).
When using these measures for ordinal classification problems, such as problems with a staging scale, the uncertainty between similar stages is quantified equally as the uncertainty between classes stage that are very far apart.
To solve this, we introduce two ordinal variants for variance and entropy that take the order of classes into account. Specifically, these measures will be higher if the uncertainty is larger between two classes that are far apart than two classes that are closer on the ordinal scale. For example, if the probabilities of class 1 and 2 are both 50\%, and so the other three classes are 0\%, the uncertainty will be lower than if the probabilities of class 1 and 5 are both 50\%. Table \ref{table:uncertainty_measures} shows how these uncertainty measure are calculated for deep ensembles.


\section{Data}
\subsection{Data}

\newcommand{\heidtest}{H\texorpdfstring{\textsubscript{test}}{Htest}}
\newcommand{\heidtrain}{H\texorpdfstring{\textsubscript{train}}{Htrain}}
\newcommand{\heidval}{H\texorpdfstring{\textsubscript{val}}{Hval}}
\newcommand{\topcon}{T\texorpdfstring{\textsubscript{test}}{Ttest}}
\newcommand{\topconblink}{T\texorpdfstring{\textsubscript{blink}}{Tblink}}
\newcommand{\bioptigen}{B\texorpdfstring{\textsubscript{test}}{Btest}}

\begin{table}[!b]
\centering
\caption{CIRCL grading system}
\begin{tabular}{@{}ll@{}}
\toprule
AMD Stage                    & Criteria                                                                                                                                       \\ \midrule
1. No AMD                    & \begin{tabular}[c]{@{}l@{}}No drusen or small, hard drusen \\only.\end{tabular} \\ \midrule
2. Early AMD                 & \begin{tabular}[c]{@{}l@{}}\textgreater{}10 small (\textless{}63 \textmu m), drusen and \\pigmentary changes or 1-15 \\ intermediate (63-124 \textmu m) drusen. \end{tabular}                       \\ \midrule
3. Intermediate AMD          & \begin{tabular}[c]{@{}l@{}}\textgreater{}15 intermediate (63-124 \textmu m)\\drusen or any large ($\geq$125 $\mu$m)\\ drusen or GA not in the central\\ circle of the ETDRS grid.\end{tabular} \\ \midrule
4. Advanced AMD: GA          & Presence of central GA. \\ \midrule
5. Advanced AMD: CNV         & \begin{tabular}[c]{@{}l@{}}Evidence of active or previous \\CNV lesion.\end{tabular} \\ \midrule
\begin{tabular}[c]{@{}l@{}}6. CNV without signs for\\AMD\end{tabular} & \begin{tabular}[c]{@{}l@{}}Chosen if CNV is present but no \\drusen of any size are present\\ within the Field 2.\end{tabular}\\
\midrule
7. Cannot grade              & \begin{tabular}[c]{@{}l@{}}Image is regarded as not \\gradable.\end{tabular}
\\ \bottomrule
\end{tabular}
\label{table:circl_protocol}
\end{table}

\begin{table}[ht]
\centering
\caption{AMD stage distribution in each dataset. The table shows the number of OCT scans.\\}
\begin{tabular}{@{}lccccc@{}}
\toprule
                  & \heidtrain & \heidval & \heidtest & \topcon & \bioptigen \\ \midrule
No AMD            & 597                                           & 216                                         & 212                                          & \multirow{2}{*}{942}                       & \multirow{2}{*}{115}                          \\
Early AMD         & 202                                           & 74                                          & 70                                           &                                            &                                               \\
Intermediate AMD  & 329                                           & 106                                         & 114                                          & 149                                        & 269                          \\
Advanced AMD: GA  & 72                                            & 29                                          & 28                                           & 37                                         &                                               \\
Advanced AMD: CNV & 737                                           & 236                                         & 256                                          & 56                                         &                                               \\ 
\bottomrule
\end{tabular}
\label{table:amd_stage_dataset}
\end{table}

Three different data sets from three different vendors were used to develop and evaluate the proposed solution: a dataset with Heidelberg OCTs served as a training set, referred to as \heidtrain, a validation set, referred to as \heidval, and internal test set, referred to as \heidtest~(Section \ref{sec:heidelberg}); and two external test sets were used to evaluate the generalizability of our models, one with Topcon scans, referred to as \topcon, (Section \ref{sec:topcon}) and one with Bioptigen scans, referred to as \bioptigen~(Section \ref{sec:bioptigen}).

\subsubsection{Heidelberg dataset (\heidtrain, \heidval~and \heidtest)}
\label{sec:heidelberg}
For development and internal testing, we used the European Genetic Database (EUGENDA), a large multi-center database for clinical and molecular analysis of AMD (\cite{Ven12a,Faus11}), containing 3,278 OCT from 1,013 patients in total. The training set \heidtrain, validation set \heidval, and test set \heidtest~contained 1937, 661 and 680 OCTs from 607 (60\%), 202 (20\%) and 204 (20\%) patients, respectively. 

Manual grading of the scans was performed by the Cologne Image Reading Center and Laboratory (CIRCL). They categorized the OCTs using the criteria described in Table \ref{table:circl_protocol}. Samples with grade 6 and 7 were excluded from this study. The number of OCTs for each of the remaining five stages is shown in Table \ref{table:amd_stage_dataset}.
The OCTs were acquired with a Spectralis HRA+OCT (Heidelberg Engineering, Heidelberg, Germany) scanner. We resampled all B-scans to the same pixel spacing of 13.9 \textmu m $\times$ 3.9 \textmu m. The number of B-scans in each OCT scan was left unchanged, which varied from 14 to 73.

\subsubsection{Topcon dataset (\topcon)}
\label{sec:topcon}

One of the external test sets was derived from the Rotterdam Study (\cite{ikra17}). This is a prospective cohort study in the city of Rotterdam, the Netherlands, that started in 1990 to investigate age-related diseases. There were in total 1184 OCT scans available from this dataset, originating from 713 patients. All OCTs were graded using the Wisconsin Age-related muclopathy grading system (WARMGS) (\cite{Klei91}) and manually harmonized to the CIRCL grading system. 
The number of OCTs for the resulting four classes is shown in Table \ref{table:amd_stage_dataset}.
The OCTs from this dataset were taken with an OCT scanner from Topcon Corp., Tokyo, Japan. Each OCT volume contained 128 B-scans. Similarly to the Heidelberg set, all B-scans were resampled to have a pixel spacing of 13.9 \textmu m $\times$ 3.9 \textmu m.

\subsubsection{Bioptigen dataset (\bioptigen)}
\label{sec:bioptigen}

The other external test set was described by \cite{Fars14}, containing normal patients and patients with intermediate AMD. For each of these subjects one OCT volume, acquired with an SD-OCT scanner from Bioptigen, Inc (Research Triangle Parc, NC), was available. The AREDS2 system \cite{Chew12} was used for grading and was harmonized to CIRCL grading system. 
The number of OCTs for these two classes is given in Table \ref{table:amd_stage_dataset}. All OCT volumes contained 100 B-scans and, all B-scans were again resampled to have a pixel spacing of 13.9 \textmu m $\times$ 3.9 \textmu m.

\section{Experimental design}
\begin{figure*}[!t]
	\centering
 	\includegraphics[width=.9\linewidth]{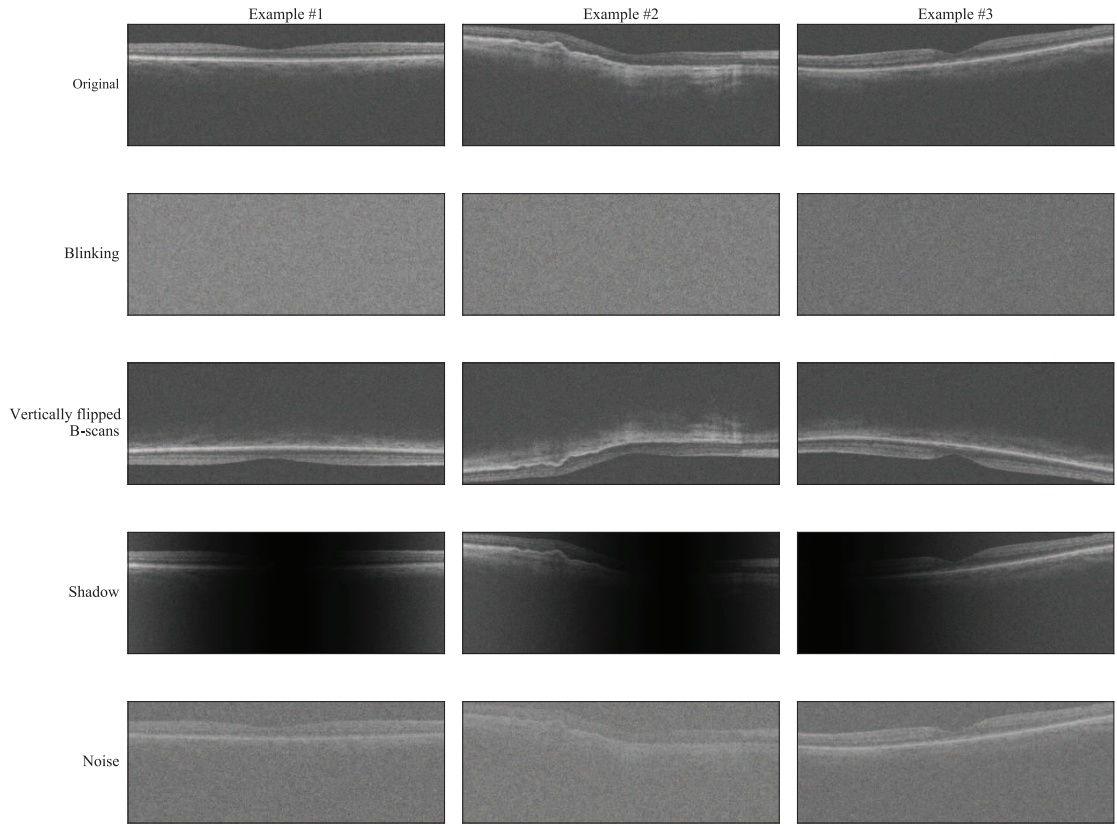}
	\caption{Examples of artificial artifacts. Each row shows the middle B-scans of a random OCT volume from \topcon. The image in the first row is the original, unaltered B-scan. The other rows each depict a different artificial artifact applied to that B-scan.}
	\label{fig:artificial_artifacts_examples}
\end{figure*}

\subsection{Vendor generalizability and interpretability}
To assess vendor generalizability of UBIX, we first calculated the performance on the internal test set, \heidtest, which is from the same distribution as the one used for training, \heidtrain. Subsequently, we evaluated the performance when transferring to the two external datasets \topcon~and \bioptigen. In the remainder of this paper, when we mention only UBIX, we refer to UBIX with the optimal $\delta$ that resulted from the grid search described later in Section \ref{sec:methods_training_network}, unless stated otherwise. We evaluated the performance on these datasets for UBIX with $\delta=\infty$ and UBIX. 
Additionally, we compared the performance of the proposed model with three different approaches, namely a 3D CNN approach, a traditional MIL approach (without UBIX) and an ensemble of multiple MIL approaches. The 3D CNN was a ResNet-18 (\cite{He15b}) with 3D convolutions and the instance-level classifiers in the MIL approaches were \mbox{ResNet-18's} with 2D convolutions.

To better show the effect of the proposed methodologies on scans with vendor-specific artifacts, we also separately evaluated the performance of the five aforementioned UBIX variants and ablations on a subset of OCT volumes in \topcon~with blinking artifacts, referred to as \topconblink. These volumes generally have multiple B-scans in which the retina is not visible. 

The interpretability of UBIX is illustrated qualitatively by showing the instance-level predictions and uncertainties for several OCT scans.

\subsection{Effect of artificial artifacts}

To demonstrate the effect of UBIX more clearly, we performed experiments where we artificially corrupted the dataset with artifacts that also occur naturally in OCT scans. The different artifact types were blinking artifacts, vertically flipped B-scans, shadows and noise. Fig. \ref{fig:artificial_artifacts_examples} shows a number of examples. We gradually introduced more OCT volumes with artificial artifacts and compared the performance for UBIX, UBIX with $\delta=\infty$ and MIL.

When one of these artifacts was applied to an OCT volume, a portion of the B-scans were affected, as happens in clinical scenarios. Artificial artifacts were then added to either one or two groups of adjacent B-scans. Both scenarios had an equal probability. The sizes of these groups had sizes of between 2\% and 15\% of B-scans, which we experimentally found to be representative of real artifacts. 

Vertically flipped B-scans are caused by a Fourier-domain detection artifact, as described by \cite{Ho10}. Shadows and noise are usually caused by media opacities, such as corneal scarring and cataract. The artificial artifacts were implemented as follows:

\begin{itemize}
    \item B-scans with artificial blinking artifacts were generated by taking an image in which all pixel values are 0 to which we applied additive random Gaussian noise for which the mean was the median pixel value in the full OCT scan and the standard deviation was equal to the standard deviation of the OCT scan. 
    \item The vertically flipped B-scans were generated by flipping the B-scans along the horizontal axis.
    \item 
To generate the shadow artifact for a particular B-scan, we adapted each A-scan (column in an OCT B-scan) $a_1,\ldots,a_A$ separately, where $A$ is the number of A-scans in the B-scan. All A-scans $a_i$ were transformed using the shadow function $S(a_i)$:

\begin{equation}
    S(a_i)=a_i(1-s(i)),
\end{equation}
where $s(i)$ is sampled from a normal probability density function:

\begin{equation}
    s(i) = \frac{1}{\sigma\sqrt{2\pi}}e^{-\frac{1}{2}(\frac{i-\mu}{\sigma})^2},
\end{equation}
where $\mu$ is randomly selected between $0$ and $A$, and $\sigma$ is randomly defined between $A/4$ and $3A/4$. $\mu$ and $\sigma$ are kept the same within one OCT volume.

\item The noise artifact is Gaussian noise added to the original image with a mean of $0$ and a standard deviation of $4$ times the standard deviation within the original OCT volume.
\end{itemize}

\subsection{Effect of uncertainty measures}
We assessed the performances of different uncertainty measures on \topcon. The uncertainty measures that we evaluated were the ones presented in Table \ref{table:uncertainty_measures}. Moreover, we separately evaluated the performance of the different uncertainty measures on \topconblink.

\subsection{Metrics}
\label{sec:methods_metrics}

For all models, to evaluate the classification performance, we calculated the area under the receiver operating characteristic curve (AUC), where intermediate and advanced AMD stages belonged to the positive class and the remaining stages to the negative class. Additionally, we computed Cohen's kappa score. For the datasets with more than two classes, the quadratic weighted kappa score ($\kappa_w$) was calculated to consider the class order. Otherwise, unweighted kappa metric was used ($\kappa$). 

To quantify artificial artifact detection performance for different uncertainty measures, we used the AUC as well, where the score was the uncertainty measure and the labels were the dichotomous variable of whether an instance had an artificial artifact or not. 
Furthermore, to estimate how well the uncertainty values were separated, we evaluated the separability of the two groups, with and without artificial artifacts, based on the uncertainty score. For this, the Xie-Beni index (XB) is calculated, defined as the ratio between cluster separation (\textit{i.e}. the minimum squared distance between cluster centers) and cluster compactness (\textit{i.e.} the mean squared distance between each data point and its cluster center (\cite{Xie91}). The lower XB, the better the data is clustered.

\subsection{Optimization and training}
\label{sec:methods_training_network}
The hyperparameters $\tau$, $\delta$ and $\gamma$ were optimized on the internal validation set \heidval. The values for $\tau$ in the grid search for UBIX with $\delta=\infty$ were $\{P_{U,80}, P_{U,80.1}, \ldots, P_{U,100}\}$, where $P_{U,i}$ is the percentile $i$ of all uncertainties in the validation set. For the UBIX grid search, the values for $\delta$ were $\{1, 5, 10, \ldots, 5\cdot10^3, 10^4\}$ and the values for $\gamma$ were $\{-0.5, -0.45, \ldots, 1.5\}$. For UBIX and ordinal entropy, performance was optimal at $\delta = 5$ and $\gamma = 1.05$. For the hyperparameter $\tau$, the optimal value was $P_{U,6.7} = 2.20$. 

The network weights were optimized with the Adam optimizer (\cite{King15}) and a learning rate of 10\textsuperscript{-4}. All images were normalized between 0 and 1. As a means of regularization, we employed online data augmentation. With a 15\% probability, random affine transformations were applied of $\pm20 \degree$ rotation within the B-scan plane, $\pm10\%$ shearing within the B-scan plane, $\pm10\%$ zooming within the B-scan plane, $\pm20$ voxels translation in the horizontal and vertical direction within the B-scan plane, $\pm2$ voxels translation in the B-scan direction. B-scans were also horizontally flipped with a 15\% probability. With a 30\% probability, random additive Gaussian noise with a mean of 0 and a standard deviation of 0.1 was applied. Also with a 30\% probability, we applied brightness modifications using the power law, varying the power between 0.75 and 3. To account for class imbalance, images were sampled during training based on their class, such that each class was sampled with an equal probability. Because of GPU memory restrictions and given that the input images were large, the batch size was 1. We used early stopping, based on $\kappa_w$ and with a patience of 10~000 batches. The deep ensembles contained 5 models, as this was shown to be sufficient for uncertainty estimation (\cite{Ovad19}).
The network weights were optimized with the Adam optimizer (\cite{King15}) and a learning rate of 10\textsuperscript{-4}. All images were normalized between 0 and 1. As a means of regularization, we employed online data augmentation. With a 15\% probability, random affine transformations were applied of $\pm20 \degree$ rotation within the B-scan plane, $\pm10\%$ shearing within the B-scan plane, $\pm10\%$ zooming within the B-scan plane, $\pm20$ voxels translation in the horizontal and vertical direction within the B-scan plane, $\pm2$ voxels translation in the B-scan direction. B-scans were also horizontally flipped with a 15\% probability. With a 30\% probability, random additive Gaussian noise with a mean of 0 and a standard deviation of 0.1 was applied. Also with a 30\% probability, we applied brightness modifications using the power law, varying the power between 0.75 and 3. To account for class imbalance, images were sampled during training based on their class, such that each class was sampled with an equal probability. Because of GPU memory restrictions and given that the input images were large, the batch size was 1. We used early stopping, based on $\kappa_w$ and with a patience of 10~000 batches. The deep ensembles contained 5 models, as this was shown to be sufficient for uncertainty estimation (\cite{Ovad19}).
The network weights were optimized with the Adam optimizer (\cite{King15}) and a learning rate of 10\textsuperscript{-4}. All images were normalized between 0 and 1. As a means of regularization, we employed online data augmentation. With a 15\% probability, random affine transformations were applied of $\pm20 \degree$ rotation within the B-scan plane, $\pm10\%$ shearing within the B-scan plane, $\pm10\%$ zooming within the B-scan plane, $\pm20$ voxels translation in the horizontal and vertical direction within the B-scan plane, $\pm2$ voxels translation in the B-scan direction. B-scans were also horizontally flipped with a 15\% probability. With a 30\% probability, random additive Gaussian noise with a mean of 0 and a standard deviation of 0.1 was applied. Also with a 30\% probability, we applied brightness modifications using the power law, varying the power between 0.75 and 3. To account for class imbalance, images were sampled during training based on their class, such that each class was sampled with an equal probability. Because of GPU memory restrictions and given that the input images were large, the batch size was 1. We used early stopping, based on $\kappa_w$ and with a patience of 10~000 batches. The deep ensembles contained 5 models, as this was shown to be sufficient for uncertainty estimation (\cite{Ovad19}).

\section{Results}
\label{sec:experiments_and_results}


\begin{table*}[!t]
\centering
\caption{Performance metrics for the different methods, evaluated on internal and external datasets. The models were all trained and validated on Heidelberg data (\heidtrain~and \heidval). The UBIX methods in this table use ordinal entropy as uncertainty measure. $n$ is the number of OCT scans. Symbols in the last two columns indicate statistically significant differences between 3D$^*$, MIL (ensemble)$^\dagger$, and MIL$^\ddagger$, calculated with non-parametric bootstrapping and $1000$ iterations (\cite{Rutt00}). We applied a Bonferroni correction to account for the number of comparisons we made.
Bolded values indicate the highest value in the row.}
\label{table:all_methods}
\begin{tabular}{@{}lllllll@{}}
\toprule
 &
  &
  3D &
  \begin{tabular}[c]{@{}l@{}}MIL\\(no ensemble)\end{tabular} &
  MIL &
  \begin{tabular}[c]{@{}l@{}}UBIX\\($\delta=\infty)$\end{tabular} &
  \begin{tabular}[c]{@{}l@{}}UBIX\end{tabular} \\ \midrule
\multirow{2}{*}{\begin{tabular}[c]{@{}l@{}}\heidtest \\ (internal, n = 680)\end{tabular}} &
$\kappa_w$ &
0.852\footnotesize{ ± 0.015}&
0.849\footnotesize{ ± 0.014}&
\textbf{0.873\footnotesize{ ± 0.013}}&
0.859\footnotesize{ ± 0.014}&
0.861\footnotesize{ ± 0.014}\\

&AUC &
0.971\footnotesize{ ± 0.006}&
0.969\footnotesize{ ± 0.006}&
0.976\footnotesize{ ± 0.005}&
0.976\footnotesize{ ± 0.005}$^\dagger$&
\textbf{0.977\footnotesize{ ± 0.005}}$^\dagger$\\

\midrule
\multirow{2}{*}{\begin{tabular}[c]{@{}l@{}}\topcon \\ (external, n = 1184)\end{tabular}} &
$\kappa_w$ &
0.341\footnotesize{ ± 0.032}&
0.601\footnotesize{ ± 0.035}&
0.624\footnotesize{ ± 0.034}&
\textbf{0.645\footnotesize{ ± 0.035}}$^*$&
0.643\footnotesize{ ± 0.036}$^*$\\

&AUC &
0.706\footnotesize{ ± 0.020}&
0.804\footnotesize{ ± 0.018}&
0.804\footnotesize{ ± 0.018}&
0.808\footnotesize{ ± 0.018}$^*$$^\ddagger$&
\textbf{0.810\footnotesize{ ± 0.018}}$^*$$^\ddagger$\\

\midrule
\multirow{2}{*}{\begin{tabular}[c]{@{}l@{}}\topconblink \\ (external, n = 33)\end{tabular}} &
$\kappa_w$ &
0.084\footnotesize{ ± 0.092}&
0.346\footnotesize{ ± 0.150}&
0.479\footnotesize{ ± 0.165}&
0.514\footnotesize{ ± 0.167}$^*$$^\dagger$&
\textbf{0.708\footnotesize{ ± 0.184}}$^*$$^\dagger$$^\ddagger$\\

&AUC &
0.622\footnotesize{ ± 0.127}&
0.727\footnotesize{ ± 0.143}&
0.872\footnotesize{ ± 0.093}&
\textbf{0.896\footnotesize{ ± 0.073}}&
0.873\footnotesize{ ± 0.093}\\

\midrule
\multirow{2}{*}{\begin{tabular}[c]{@{}l@{}}\bioptigen \\ (external, n = 384)\end{tabular}} &
$\kappa$ &
0.000\footnotesize{ ± 0.000}&
0.714\footnotesize{ ± 0.039}&
0.757\footnotesize{ ± 0.037}&
\textbf{0.806\footnotesize{ ± 0.033}}$^*$$^\dagger$&
0.783\footnotesize{ ± 0.034}$^*$\\

&AUC &
0.702\footnotesize{ ± 0.028}&
0.914\footnotesize{ ± 0.018}&
0.915\footnotesize{ ± 0.020}&
\textbf{0.924\footnotesize{ ± 0.020}}$^*$$^\ddagger$&
0.922\footnotesize{ ± 0.020}$^*$\\

\bottomrule

\end{tabular}
\end{table*}

\begin{figure*}[!t]
        \centering
        \includegraphics[width=.95\textwidth]{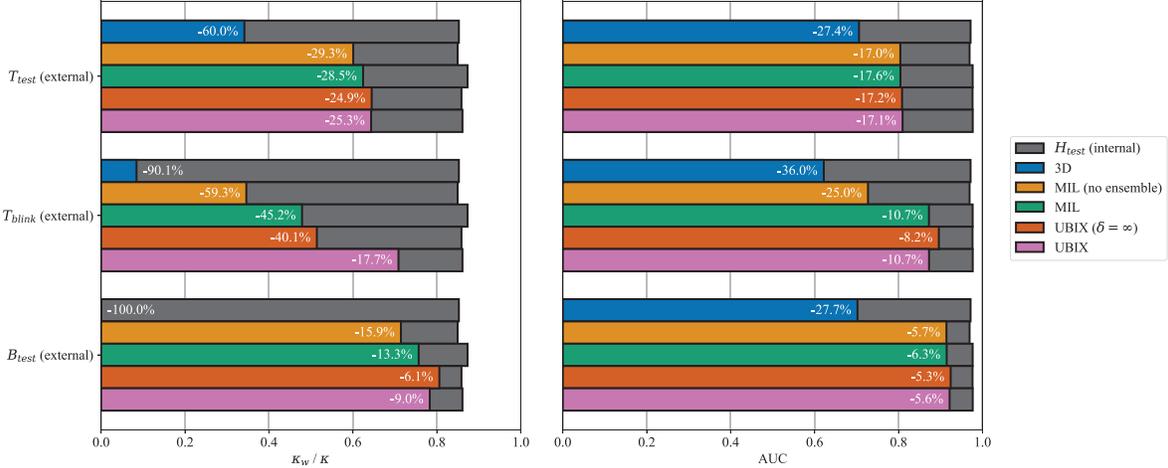}
	\caption{Performance differences between datasets of the various methods. In the left plot, $\kappa_w$ is shown for \heidtest, \topcon~and \topconblink, and $\kappa$ is shown for \bioptigen. In the right plot, AUC is shown. The absolute performances are indicated by the width of the bar plots, while the performance differences of the three external datasets compared to the internal test set are indicated with text.}
	\label{fig:performance_drops}
\end{figure*}

\begin{figure*}[!t]
	\centering
    
    \begin{subfigure}[]{0.95\textwidth}
        \centering
        \includegraphics[width=\textwidth]{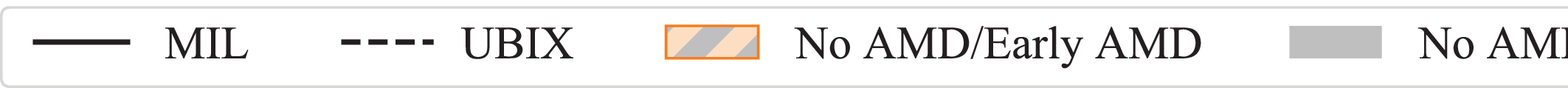}
    \end{subfigure}

    \begin{subfigure}[]{0.8\textwidth}
        \centering
        \includegraphics[width=\textwidth]{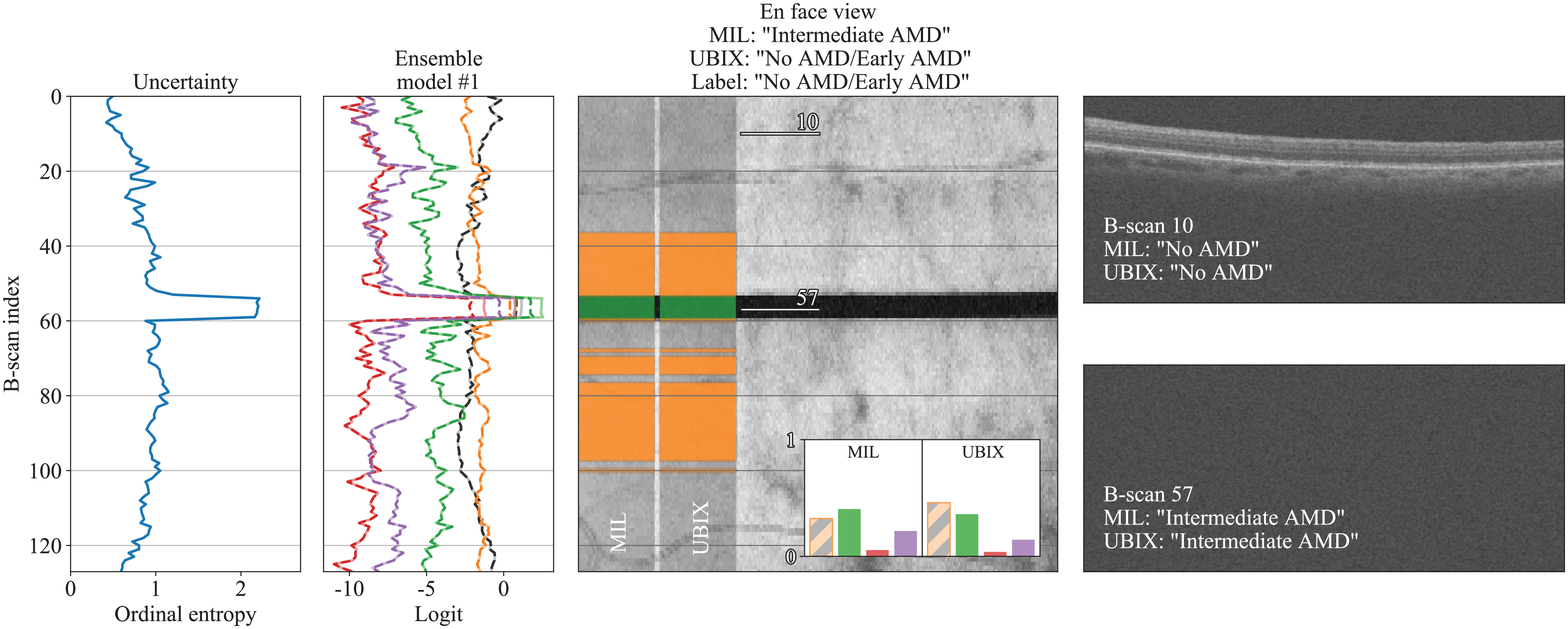}
        \caption{UBIX correctly predicts "No AMD/Early AMD", while MIL incorrectly predicts "Intermediate AMD". UBIX suppresses the instance-level outputs at the location of the artifact around B-scan 57, causing it to be robust to that artifact, in contrast to MIL.}
    \end{subfigure}
    \begin{subfigure}[]{0.8\textwidth}
        \centering
        \includegraphics[width=\textwidth]{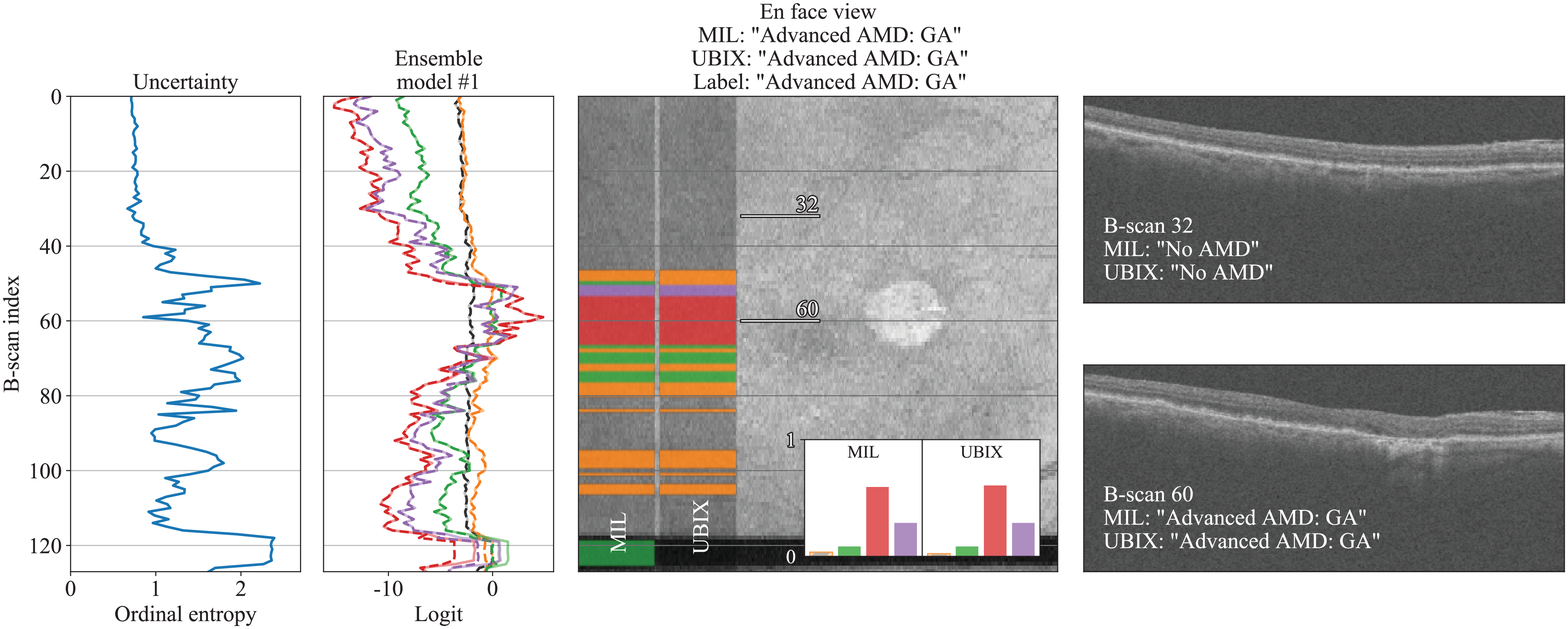}
        \caption{Correct GA volume-level classification. Central GA is also well visible in the en face image as a white circular object around B-scan 60, which is pointed out by the instance-level output in red.}
    \end{subfigure}
	\caption{Examples where UBIX corrects volume-level and instance-level predictions. The figure also illustrates instance-level interpretability. Each subfigure shows, from left to right, the uncertainty per instance, the instance-level logits of the first model in the ensemble (only showing one model for clarity), the en face image (the volume averaged over the y-axis), and two B-scans of interest. The uncertainty, logit and en face plots correspond spatially in the horizontal direction. The left and right banner in the en face view indicate the instance-level outputs for MIL and UBIX, respectively. The B-scans on the right are highlighted in the en face view. In the bottom right of the en face views, volume-level probabilities are shown.}
	\label{fig:interpretability_examples}
\end{figure*}

Table \ref{table:all_methods} shows the performance of the two UBIX variants, compared to three ablated models. Fig. \ref{fig:performance_drops} shows the performance differences in percentages for these methods. On the internal dataset, most differences between models are smaller than those found for the external datasets. 
The performance on the external dataset drops compared to the internal dataset for all methods. However, in terms of $\kappa_w$ this drop is always smaller for the UBIX models than for the other models. 
UBIX especially outperforms the ablated models on cases with artifacts; on \topconblink, $\kappa_w$ is $0.708$ for UBIX, while it was $0.479$ for only MIL, $0.346$ for MIL (no ensemble), and $0.084$ for 3D. 

Fig. \ref{fig:interpretability_examples} shows a number of visual examples of UBIX performance. The figure also illustrates the interpretability of the model at B-scan level.

\begin{figure*}[!t]
	\centering
	\includegraphics[width=.9\linewidth]{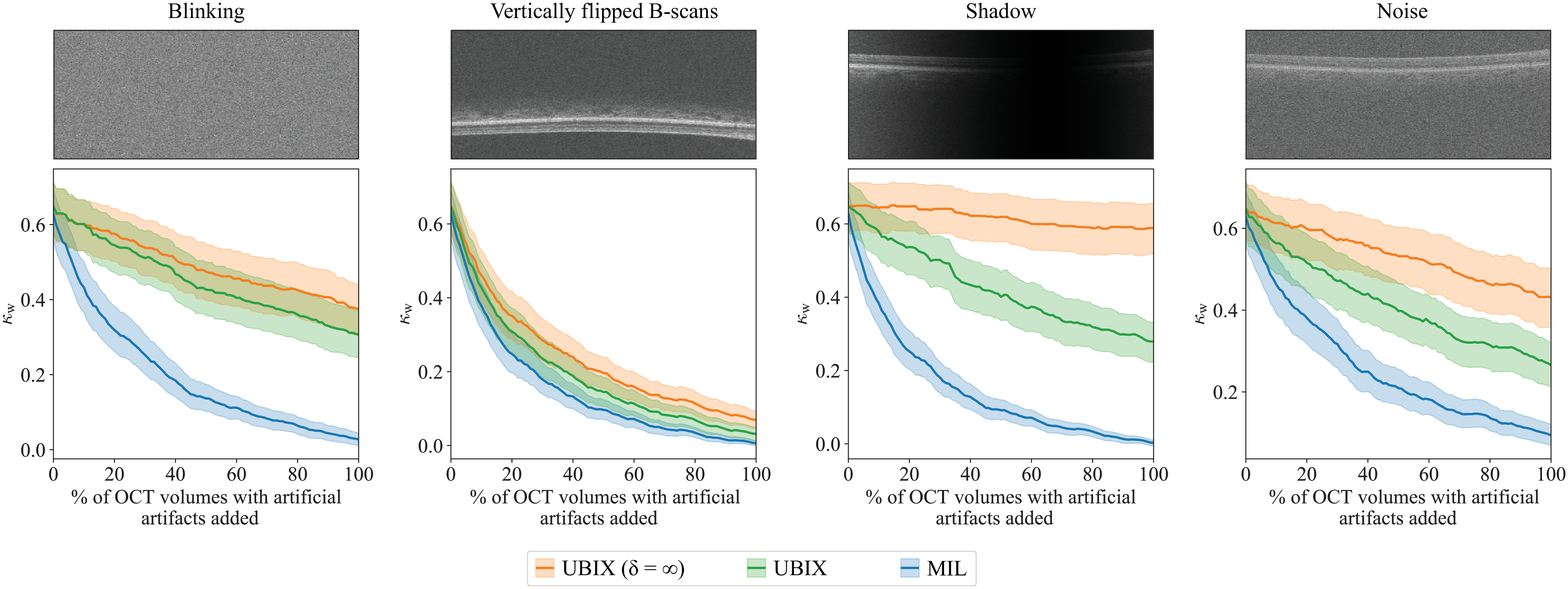}
	\caption{Robustness to different artificial artifacts of the UBIX, compared to MIL, on \topcon. The top image in each column shows an example B-scan of the artificial artifact. The plots show the relation between the performance and the percentage of OCT volumes in the dataset that contain these artifacts. The shaded areas indicate 95\% confidence intervals, obtained using bootstrapping with 1000 iterations.}
	\label{fig:artificial_artifacts_robustness_results}
\end{figure*}

Fig. \ref{fig:artificial_artifacts_robustness_results} shows the effect of adding artificial artifacts to the dataset when using UBIX, compared to MIL, by evaluating their performance when varying the percentage of OCT volumes that are affected by artificial artifacts in the \topcon~dataset.

Table \ref{table:uncertainty_measures_results} shows the performances on \topcon~ of UBIX when using different uncertainty measures, where the hyperparameters were optimized separately on the internal validation set for each uncertainty measure. For \topcon, $\kappa_w$ is highest for ordinal entropy.

\begin{figure*}[!t]
	\centering
	\includegraphics[width=.8\linewidth]{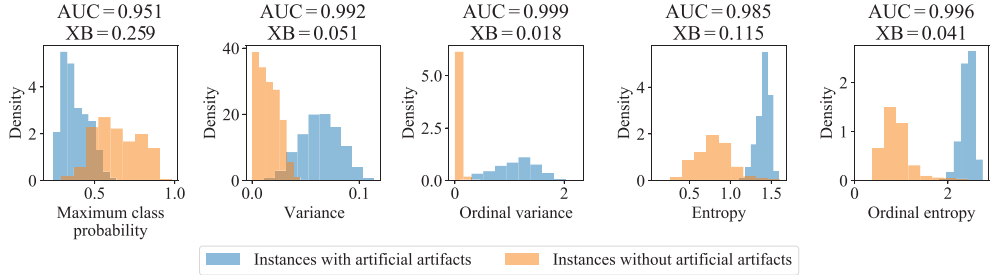}
	\caption{Density plots of uncertainty estimates at instances with artificial blinking artifacts compared to uncertainty estimates at instances without artificial artifacts, applied to \topcon. The AUC for artificial artifact detection performance and the clustering metric XB (the lower, the better) are shown above the density plots.}  
	\label{fig:uncertainy_measures_at_artificial_artifacts}
\end{figure*}

Fig. \ref{fig:uncertainy_measures_at_artificial_artifacts} visualizes the different uncertainty measures for artificially added artifacts. The XB value and AUC are largest when using ordinal variance or ordinal entropy.

\begin{table*}[ht]
\centering
\caption{Performance metrics for different uncertainty measures used the during instance exclusion step of the UBIX approach, evaluated on \topcon. The models were all trained and validated on Heidelberg data (\heidtrain~and \heidval). $n$ is the number of OCT scans. Bolded values indicate the highest value in the row.}
\label{table:uncertainty_measures_results}
\begin{tabular}{@{}lllllll@{}}
\toprule
 &
   &
  \begin{tabular}[c]{@{}l@{}}Maximum class\\ probability\end{tabular} &
  \begin{tabular}[c]{@{}l@{}}Mean class\\ variance\end{tabular} &
  \begin{tabular}[c]{@{}l@{}}Ordinal\\ variance\end{tabular} &
  Entropy &
  \begin{tabular}[c]{@{}l@{}}Ordinal\\ entropy\end{tabular} \\ \midrule
\multirow{2}{*}{\begin{tabular}[c]{@{}l@{}}\topcon \\ (external, n = 1184)\end{tabular}} &
$\kappa_w$ &
0.605\footnotesize{ ± 0.037}&
0.583\footnotesize{ ± 0.038}&
0.615\footnotesize{ ± 0.035}&
0.617\footnotesize{ ± 0.038}&
\textbf{0.643\footnotesize{ ± 0.036}}\\

&AUC &
0.802\footnotesize{ ± 0.018}&
0.804\footnotesize{ ± 0.018}&
0.805\footnotesize{ ± 0.018}&
\textbf{0.812\footnotesize{ ± 0.018}}&
0.810\footnotesize{ ± 0.018}\\

\midrule
\multirow{2}{*}{\begin{tabular}[c]{@{}l@{}}\topconblink \\ (external, n = 33)\end{tabular}} &
$\kappa_w$ &
0.497\footnotesize{ ± 0.167}&
0.686\footnotesize{ ± 0.186}&
0.498\footnotesize{ ± 0.166}&
\textbf{0.732\footnotesize{ ± 0.186}}&
0.708\footnotesize{ ± 0.184}\\

&AUC &
0.772\footnotesize{ ± 0.117}&
\textbf{0.890\footnotesize{ ± 0.078}}&
0.879\footnotesize{ ± 0.087}&
0.831\footnotesize{ ± 0.130}&
0.873\footnotesize{ ± 0.093}\\

\bottomrule

\end{tabular}
\end{table*}

\section{Discussion}
\label{sec:discussion}

We proposed a method using MIL with OOD detection to improve the generalizability of deep learning models for classification of 3D medical images. The model aims to reduce the effect of on-the-fly detected OOD instances in the final classification of the bag.
By suppressing the contribution of OOD instances, our proposed UBIX function maintains performance on unseen data distributions, namely images coming from different scanners. 

The robustness of the proposed approach was demonstrated by transferring UBIX models and corresponding ablated versions to external datasets from different vendors. As shown in Fig. \ref{fig:performance_drops}, UBIX variants are less prone to significant performance drops than the other models. On all external datasets, either UBIX or UBIX with $\delta = \infty$ showed better results than the ablated models in terms of absolute performance. The performance drop was most notable on \topconblink, where UBIX maintained a $\kappa_w$ of 0.708, while the best and worst performing ablation models (MIL and 3D, respectively) has a $\kappa_w$ of 0.479 and Kw of 0.084, respectively. 
It was expected that these performance differences were more notable on \topconblink, which only contained OCTs with blinking artifacts, because UBIX was designed to be robust to vendor-specific artifacts.

It was noted that, depending on the inference data, it was preferable to fully exclude the outputs of uncertain instances (UBIX with $\delta = \infty$) or to only suppressed them (UBIX). From Fig. \ref{fig:artificial_artifacts_robustness_results}, we found that UBIX with $\delta=\infty$ showed better robustness than UBIX for OCTs with artificial artifacts. A possible reason for this could be that some of the artificial artifacts highly corrupted the information, resulting in a notably strong incorrect signal and the requirement for full exclusion of uncertain instances. UBIX with $\delta=\infty$ seemed to be especially robust to shadow artifacts, given that the performance barely decreased when introducing more OCT volumes with artificial artifacts. For some external datasets and metrics, however, UBIX achieved better results than UBIX with $\delta=\infty$, e.g. for $\kappa_w$ on \topconblink~and AUC on \topcon. 

The performed data augmentation might also have an effect on generalizability. Since signal-to-noise ratios differ per scanner, noise augmentation probably aided our models to generalize. Although we did resample all images to have the same pixel spacing within B-scans, the original spacings that we had could have been slightly inaccurate. Therefore, zooming could potentially also have a positive effect on generalizability. The same type of data augmentation was applied in all experiments and measuring its effect was considered out of the scope for this paper. 


Large variability in B-scan spacing between scanners can also cause features learned by 3D CNNs to be poorly generalizable. MIL, which processes B-scans individually and combines B-scan level outputs using a MIL pooling function to get a volume level output, improves the robustness to this variability in slice spacing with respect to 3D models. 
We observed that performance differences are minimal between a 3D CNN and MIL when evaluated on data from the same vendor used during training. When evaluating on data from a different vendor, a performance drop was observed for both methods, although this drop is much larger for the 3D method (60.0\% drop in $\kappa_w$ on \topcon) than for MIL (29.3\% in $\kappa_w$ on \topcon).

The extent to which UBIX improves generalizability depends highly on the quality of its underlying OOD detection. Therefore, we compared three different commonly used uncertainty measures, and we proposed two ordinal variants. On \topcon, entropy and its ordinal variant had the highest performance in terms of $\kappa_w$ and AUC, respectively. The ordinal variants seemed to distinguish the B-scans with and without artificial artifacts the best. This can be seen in the density plots of Fig. \ref{fig:uncertainy_measures_at_artificial_artifacts}, and this is also reflected in the AUC and XB values, which were best for the two ordinal variants. 
Hence, the ordinal variants led to higher performances on artificial artifacts, although in \topcon~we find mixed observations (Table \ref{table:uncertainty_measures_results}). A possible reason for this could be that fewer evaluation data with real artifacts were available, resulting in a less accurate performance measurement than when using artificial artifacts.

One of the advantages of using MIL as base of the UBIX model is that instance-level annotations are not required for training, while the model is able to produce a classification output at this level (B-scan level in our case) as well as calculating instance-level uncertainty.
This introduces model explainability, increasing the transparency of our method, and allowing surveillance of its behaviour.

Since the three datasets were acquired and annotated at different sites with varying protocols, the reference standards were set different among these datasets. 
To minimize the effect of this discrepancy, we merged the first two classes of the CIRCL grading when evaluating with the WARMGS system which was available for \topcon. Moreover, when evaluating on \bioptigen~for which only the binary labels \textit{No AMD} and \textit{Intermediate AMD} were available, we also binarized the CIRCL systems, where the positive class started at \textit{Intermediate AMD}. This harmonization approach, however, was not perfect, causing the resulting class definitions to still not be completely equal. Despite this discrepancy, we think measuring performance differences between methods is still well possible. Nevertheless, the absolute performances can be underestimated because of these differences in reference standards.

As a potential undesired side effect, it should be noted that difficult cases, which are assumed be more uncertain, could be excluded which are in fact necessary for making a correct prediction. It will depend on the setting of model deployment whether the benefits of robustness to OOD data outweigh this drawback. In screening, for example, a high specificity is generally considered more important than a high sensitivity. Instances that are falsely excluded by UBIX will often contain abnormalities, in which case especially sensitivity would suffer, but specificity will not be affected in that case. When the task is disease staging (as is the case in this work) and UBIX is implemented with an ordinal uncertainty measure, this potential undesired side-effect is unlikely to occur. Difficulty in such a case will usually lie in the uncertainty between two classes that are close together in the staging scale (such as \textit{Early AMD} and \textit{Intermediate AMD}), resulting in a low estimate of the ordinal uncertainty measure. As artifacts are not likely to cause uncertainties between two classes that are close on the staging scale, these artifacts will probably have a much higher ordinal uncertainty measure. To give an indication of which instances were assigned the highest uncertainties, we manually analyzed the B-scans with the highest ordinal entropies in \topcon~in Section A of the Appendix. There we found that more than half of the B-scans in the first percentile of most uncertain ones contained one of the nine different types of artifacts that were seemingly related to image acquisition.

If relevant structures are entirely excluded because they are difficult for the model to classify (for example because of an unseen lesion type or ambiguity), this is likely to be because there is something atypical with the whole scan or setting in which the model is used and the user should be alerted. So instead of only silently excluding instances, future work could analyze a method for combining UBIX with alerting the user if there are too many OOD instances detected. 
In future work, UBIX could also be adapted to work with patches as instances instead of slices. If an artifact is only locally present within a slice, the entire slice would not be excluded and potentially useful information would not be ignored.

We did not compare our method to any other domain adaptation methods which often require additional supervised or unsupervised training. Such a comparison would be unfair, as our method does not require any additional training. Nevertheless, our approach could potentially be further improved with the incorporation of domain adaptation methods such as those proposed by \cite{Seeb19} and \cite{Romo20}.
Future work could also investigate different OOD detection methods to be incorporated in UBIX, as UBIX is theoretically compatible with any OOD detection method. Performing a systematic analysis for OOD detection methods was beyond the scope of this paper, so we only applied deep ensembles in this study. Such a comparison might lead to valuable insights and performance improvements.
In this work, we only evaluated our approach for AMD grading in OCT. The method is expected to be applicable in more problem settings, such as the classification of other features and retinal diseases in OCT, but potentially also in other medical image analysis applications.






\section{Conclusion}
\label{sec:conclusion}

We showed that the generalizability of classification models to unseen scenarios can be improved by UBIX, an approach that seamlessly suppresses the contribution of OOD instances to the final classification during test-time based on the uncertainty associated with these instances, in the context of MIL for AMD classification in OCT. Our proposed approach alleviates the need for retraining on new data, which is an expensive process in terms of data acquisition, model development, and human annotation time. This increases reliability by improving the applicability of artificial intelligence models for OCT classification in broader scopes than the settings in which they were developed.

\section*{Acknowledgement}
This research was funded by Eurostars grant E12712.

\bibliographystyle{unsrt}  
\bibliography{bib/fullstrings,bib/diag,bib/diagnoweb,bib/newrefs}

\end{document}